\documentclass{elsart5p}

\usepackage{graphicx}

\usepackage{amssymb}
\newcommand {\tbti}{Tb$_{2}$Ti$_{2}$O$_7$}

\newlength{\figwidth}

\newcommand{\Rti}{R$_{2}$Ti$_{2}$O$_7$}
\newcommand{\sinth}{\ensuremath{\sin\theta/\lambda}}

\begin{document}

\begin{frontmatter}



\title{Spin Density and Non-Collinear Magnetization in Frustrated Pyrochlore \tbti\ from Polarized Neutron Scattering.}


\author{{Arsen Gukasov$^1$, Huibo Cao$^1$, Isabelle Mirebeau$^1$  and Pierre Bonville$^2$}}

\address{$^1$Laboratoire L\'eon Brillouin, CEA-CNRS, CE-Saclay, 91191
Gif-sur-Yvette, France.}
\address{$^2$Service de Physique de l'Etat Condens\'e,
CEA-CNRS, CE-Saclay,  91191 Gif-Sur-Yvette, France.}
\begin{abstract}{We used a local susceptibility  approach in extensive polarized neutron diffraction studies of the spin liquid \tbti. For a magnetic field applied along the [110] and [111] directions, we found that, at high temperature, all Tb moments are collinear and parallel to the field. With decreasing temperature, the Tb moments reorient from the field direction to their local anisotropy axes. For the [110] field direction, the field induced magnetic structure at 10 K is spin ice-like, but with two types of Tb moments of very different magnitudes. For a field along [111], the magnetic structure resembles the so-called "one in-three out" found in spin ices, with the difference that all Tb moments have an additional component along the [111] direction due to the magnetic field.
The temperature evolution of the local susceptibilities clearly demonstrates  a progressive change from Heisenberg to Ising behavior of the Tb moments when lowering the temperature, which appears to be a crystal field effect.}
\end{abstract}

\end{frontmatter}

\section{Introduction}
Recent polarized neutron studies of the magnetic distribution in some ferromagnetic and paramagnetic materials with the $Th_3P_4$ cubic structure have
shown that the moment induced by a magnetic field at equivalent crystallographic sites may be very different \cite{Gukasov,SmTe,SmTe2}. Such a behavior can arise when the local symmetry of the magnetic ion is lower than the overall symmetry of the crystal. The moment induced on each magnetic ion by the internal or external magnetic field depends on the orientation of the field with respect to the local symmetry axis.
This effect can be described by attributing to each magnetic atom a
site susceptibility tensor $\chi_{ij}$ which gives the magnetic (linear)
response of the ion to the applied magnetic field \cite{gukasov-brown}.  The symmetry of the $\chi_{ij}$ tensor is the same as that
of the $u_{ij}$ tensor  describing the thermal motion of atoms.  The
components of the $u_{ij}$ tensor represent the mean square atomic
displacement parameters (ADPs). In the linear approximation, i.e. for relatively weak magnetic fields, one can
introduce atomic suceptibility parameters (ASPs) in a way analogous to the
ADPs. The response of an
atom to a magnetic field can then be conveniently visualized as a {\em
magnetization ellipsoid} constructed from the six independent ASPs in
much the same way as {\em thermal ellipsoids} are constructed from the
ADPs. In the absence of local anisotropy the  magnetic
ellipsoids reduce to spheres with their radii proportional to the
induced magnetization, but in many cases {\em anomalous} (elongated or
flattened) ellipsoids can appear. The presence of such anisotropic
magnetic ellipsoids  accounts for the anomalous magnetic behavior mentioned above.

\Rti\ pyrochlore compounds represent another type of a cubic structure with space group {\it Fd $\bar{3}$m} where the local symmetry of the magnetic ion is lower than the overall symmetry of the crystal.  In \Rti, the rare earth ions
occupy a 16d site with a local trigonal symmetry $\bar{3}m$ and are subject to a strong  crystal field interaction. Both the R
and Ti atoms independently lie on a pyrochlore
structure, a face-centered cubic lattice of corner sharing
tetrahedra which produces phenomena
of geometrical frustration. Depending on the
nature of the magnetic rare earth ion in these materials,
the ground state can exhibit long range magnetic
order \cite{Champion03,Raju99} or spin ice physics \cite{Harris97,Hertog00}, and in the case of
\tbti it is a highly correlated quantum disordered state
known as a  spin liquid \cite{Gardner99,Gardner01,Gardner03}.
In \tbti\ \cite{Gingras00,Mirebeau07} the anisotropy is weaker than in canonical spin
ices, and ferromagnetic (F) and antiferromagnetic (AF) first
neighbor interactions nearly compensate. This makes \tbti\ extremely sensitive to external perturbations, which leads to a rich
variety of magnetic ground states, from spin liquid \cite{Gardner99} to antiferromagnetic order under
pressure \cite{Mirebeau02} and/or in magnetic field \cite{Mirebeau04,Rule06}.
We undertook a systematic study of the field induced magnetic orders in the spin liquid \tbti\ in a wide temperature ($0.3<T<270$ K) and field $(0<H<7$ T) range, by combining polarized and unpolarized neutron diffraction on a
single crystal. Unpolarized neutron results will be  published elsewhere \cite{Cao08,CaoHFM}; we describe here in detail our  polarized neutron results based on the local susceptibility approach.
We show that this approach  is extremely efficient in the treatment of  polarized neutron  data  as it allows an universal description of magnetic structures existing in  large temperature (5-300K) and field (1-7 T) ranges, regardless of the field direction, with only 2 parameters.

\section{Experimental}
Neutron diffraction studies were performed on the
diffractometer Super-6T2 \cite{6t2} and 5C1 at  the ORPH\'EE reactor of the L\'eon Brillouin Laboratory, CEA/CNRS Saclay.
Polarized neutron flipping ratios  were measured on Super-6T2 using neutrons with
$\lambda_{n}=1.4$~\AA ~obtained with a supermirror bender   ($P_0=0.98$) and completed by measurements on 5C1
($\lambda_{n}=0.84$~\AA) ~obtained with a Heussler polarizer  ($P_0=0.91$).
The programs  CHILSQ of the Cambridge Crystallography
Subroutine Library \cite{ccsl} and MEND \cite{takata} were used for
the least squares
refinements on the  flipping ratios and maximum entropy reconstruction respectively.

A single crystal of  \tbti\  was grown from a sintered rod of the same nominal composition
by the floating-zone technique, using a mirror furnace \cite{Cao08}.
Prior to polarized neutron
measurements  the crystal was characterized by neutron diffraction at 180 K and 2 K in zero-field.
A total of 238  reflections
with $\sinth\ < 0.6$ \AA $^{-1}$ was measured at 2~K. The structure factors of 46  unique reflections were
obtained by averaging equivalents. These were used to refine all
positional parameters, the occupancy factors, the isotropic
temperature factors and the extinction parameters.
The refinement of the crystal structure was performed using   space group {\it Fd$\bar{3}$m}.

\section{Magnetic ellipsoids and anisotropic susceptibility parameters}
According to Ref. \cite{gukasov-brown} atomic (local) susceptibility parameters can be determined from polarized neutron flipping ratio
measurements if the proper magnetic symmetry of the crystal is taken
into account.
In order to investigate the temperature dependence of local susceptibilities of \tbti, a series of
 flipping ratio measurements was made at 1.6, 5, 10, 20, 50, 100, 150, 200 and 270 K with a magnetic
field of 1 T applied  parallel to the [110]
direction. (In fact there was a
misalignment of about 5 degrees between the [110] direction  and the magnetic
field direction. The  exact orientation of the magnetic field
 with respect to the cubic crystal axes was taken properly
into account in the final data analysis.)  The results were interpreted in
terms of a model which assigns a
site susceptibility tensor {\boldmath$\chi$} to each
crystallographically independent site \cite{gukasov-brown}.
The point group symmetries of the
atomic sites of the paramagnetic group {\it Fd$\bar{3}$m} were used to
constrain the site susceptibility tensors.
In the cubic axes, the symmetry constraints for a magnetic
atom occupying the 16d site in the {\it Fd$\bar{3}$m} group imply:
$\chi_{11}=\chi_{22}=\chi_{33}$ and  $\chi_{12}=\chi_{13}=\chi_{23}$.
Thus, only two independent susceptibility parameters $\chi_{11}$ and  $\chi_{12}$ need to be
determined regardless of the field direction.

For each temperature these two independent components $\chi_{11}$ and $\chi_{12}$ were determined using the least squares refinement program CHILSQ
\cite{ccsl}.
The  refinement carried out with the 102 flipping ratios
measured at 270 K with the magnetic field $H=1$~T applied parallel to [110]
gave $\chi_{11}=0.056(5)$ $\mu_B/$T and $\chi_{12}=
0.005(6)(11)$ $\mu_B/$T with a goodness of fit $\chi^2$=1.02. For
convenience, the susceptibility components are given in $\mu_B$/T, which allows an easy comparison with the results of a conventional least
squares refinement based on the localized magnetic moment model. The fact that the non-diagonal term
$\chi_{12}$ is equal to zero (within  error bars) shows that the  magnetic ellipsoids at
270 K reduce to spheres of diameter 0.056 $\mu_B/$T. It means that the moments induced on all Tb atoms in the unit cell
 have the same magnitude and are collinear with the magnetic field regardless of the field direction. This behavior resembles that of a Heisenberg  paramagnet at high temperature.
The situation is quite different at low temperature where
the refinement carried out with the 423 flipping ratios
measured at 10 K in the same magnetic field
gave $\chi_{11}=0.939(15)$ $\mu_B/$T and $\chi_{12}=
0.535(11)$ $\mu_B/$T with the goodness of fit $\chi^2$=2.92 .
One can see that, besides a considerable increase of the diagonal term $\chi_{11}$ reflecting the  Curie-Weiss behavior, the non-diagonal elements $\chi_{12}$ become very large, showing that
the principal axes of the magnetic ellipsoid are different in length and do not coincide
with the cubic crystal axes. It is easy to show that the trigonal local symmetry implies that the
main principal axes of the ellipsoids lie along the four [111]
axes and their lengths are given by: \newline  $\chi_\parallel$=$\chi_{11}$+2$\chi_{12}$ and $\chi_\perp$=$\chi_{11}$-$\chi_{12}$.
\begin{figure*}[ht]
\begin{center}
\includegraphics[width=35pc]{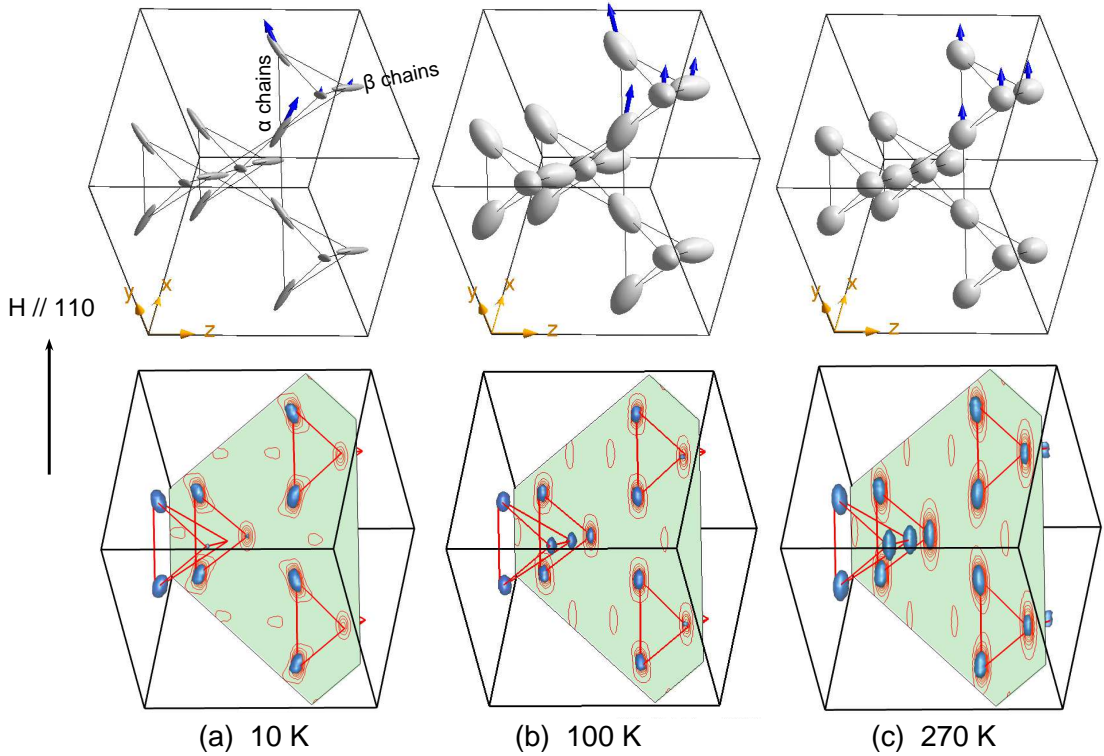}
\end{center}
\caption{\label{fig1} Top: Magnetic structure (arrows) and magnetic
ellipsoids measured in the field  H=1 T $\parallel$ to [110] at 10 K (a), 100 K (b) and 270 K (c). Ellipsoids were scaled by temperature to compensate the Curie-Weiss behavior.
Bottom: Two dimensional (2D) sections of the  magnetization density component $m_Z$ parallel to the field (red contours) and three dimensional (3D) isosurfaces of the magnetization density $m_Z$ (blue). The level of the isosurface selected on each figure  corresponds to 25\% of the maximal density in the cell found in the reconstruction. The maximum values of densities are scaled by temperature.}
\end{figure*}

The evolution of the magnetic ellipsoids with temperature  is
illustrated in Fig.\ref{fig1}.
One can see  that the shape and
the mutual orientation of the ellipsoids reflect the cubic
symmetry of the structure. The ratio of the length of the main
axes, $\chi_\parallel$ along the local anisotropy (trigonal) axis [111]  and $\chi_\perp$ perpendicular
to it, is strongly  temperature dependent and the ellipsoid elongation, which is due to an enhancement of
the  $\chi_{12}$  component,
increases markedly at low temperatures.
Magnetization along the easy and hard local axes at 10 K  differ
by a factor of 5, which shows that the local anisotropy progressively evolves  from Heisenberg to Ising character when the temperature decreases.

Once the site susceptibility parameters are known, one can easily
calculate the magnitude and the direction of the moment induced on
each Tb atom  by a magnetic field applied in an
arbitrary direction
since the site susceptibility tensor relates the vectors {\bf M} on each Tb ion
and {\bf H}: {\bf M} = {\boldmath$\chi$} {\bf H}. The  magnetic structures at 10, 100 and 270 K resulting from the refinement  are shown in Fig.\ref{fig1}. There is a progressive evolution of the structure from collinear, with the same magnetic moment on all Tb ions, to non-collinear with two types of Tb ions having a high and a low moment.
This corresponds to the separation of Tb ions into so-called  $\alpha$ and $\beta$-chains \cite{Hiroi03,Ruff05}. Moments in $\alpha$-chains have their local $<111>$ easy axis close to {\bf H}
(35.3$^\circ$), whereas moments in $\beta$-chains have their easy
axis perpendicular to {\bf H}.
 The non-collinear magnetic structure induced at low temperature by the field can be seen as
a direct consequence of the strong local anisotropy, the moments remaining close to the elongated axes of the ellipsoids whatever the field direction.

\begin{figure}[ht]
\begin{center}
\includegraphics[width=20pc]{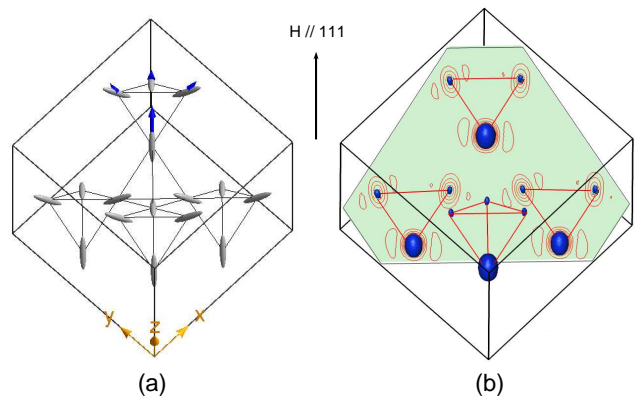}
\end{center}
\caption{\label{fig2} a) Magnetic structure  and magnetic
ellipsoids measured in a field  H=1 T $\parallel$ to [111] at 10 K.
b) The corresponding 2D map and 3D isosurface of the magnetization density $m_Z$, see captions to Fig.~1.}
\end{figure}

To verify the validity of the local susceptibility approach, a second series of flipping ratio measurements
was performed with a field of 1 T parallel to the [111] direction at different temperatures.
Refinement on 144 flipping ratios at 10 K  gave
$\chi_{11}= 0.95(5)$ $\mu_B$ and $\chi_{12}= 0.55(4)$
$\mu_B$ with $\chi^2$=7.1. The $\chi_{ij}$ parameters for the [111] field direction coincide within
experimental error with those obtained for the field along [110].
The  magnetic structure at 10 K resulting from the refinement is shown in
Fig.\ref{fig2}.
One can easily recognize the so-called "one in-three out" spin configuration described earlier in Ref.\cite{Hiroi03}, with the difference that all Tb moments have additional components along the [111] direction due to the magnetic field.
These results confirm that the polarized neutron
data collected with different field orientations can be interpreted
within the paramagnetic group {\it Fd$\bar{3}$m} using only the 2
susceptibility parameters allowed by symmetry.

\section{Magnetization densities}
Our refinements show a significant difference between the magnetic moments in  $\alpha$ and $\beta$-chains  and it is useful to check to what extent this result can be substantiated by a model-free analysis
of our data by the maximum
entropy method (MEM). This method has been shown to give much more reliable results than
conventional Fourier syntheses, by considerably reducing both noise and truncation effects \cite{papou}.
In order to carry out the MEM reconstructions the magnetic structure factors of all data sets
were extracted  from flipping ratios using the SORGAM program \cite{ccsl}.

 As shown in Ref. \cite{SmTe} when the local symmetry of the magnetic ion is lower than the overall symmetry of the crystal, the high symmetry paramagnetic group {\it Fd$\bar{3}$m} cannot be used in the magnetization density reconstruction. For the field oriented along
the two-fold axis [110], for example, the low symmetry (orthorhombic)
$Fdd2$ group must be used.
This is the highest symmetry subgroup of
{\it Fd$\bar{3}$m} under which the homogeneous magnetization component
$m_Z$, induced parallel to the two-fold axis by the magnetic field,
is invariant. An important  consequence of the symmetry reductions is that the number of independent  reflections to be measured  increases considerably  compared to that of the paramagnetic group.  A typical number of  reflections used in the reconstruction  varies from 100 to 400, depending on the degree of local anisotropy and on the magnitude of the flipping ratios. The magnetization density distribution was discretized
into 51$\times$51$\times$51 sections along a, b and c respectively. Then the magnetization density component parallel to the field ($m_Z$) was reconstructed using a conventional {\it uniform} (flat) density \cite{papou}.
Results of the reconstruction are presented in Fig.\ref{fig1} a-c. For each temperature, the figure shows both a 2D section of the magnetization density in the [111] plane and a 3D isosurface of the density. One can see that at 270 K  the magnetization isosurface is nearly the same on $\alpha$ and $\beta$-chains, while at 10 K magnetization is practically absent on the $\beta$-chains.
Thus the  reconstruction confirms the presence of a large difference of $m_Z$ components of the Tb moment in $\alpha$ and $\beta$-chains at low temperature, in spite of the fact that the maximum entropy method
is strongly biased against a disequilibrium of magnetic density in the unit cell.
 We note also that sections through a Tb site at 10 K show that the
 magnetization density on $\alpha$ -chains is elongated in the [111] directions (Fig.\ref{fig1} a, bottom). This might be
 a consequence of the strong crystal field  anisotropy at low temperature. We stress that, in the above figures, we only present the $m_Z$ component of the density as the
 flipping ratio measurements can only yield the density component collinear with the field.
On the other hand the presence of a high degree of anisotropy in the local susceptibility implies a considerable non-collinearity on the magnetization distribution at low temperature. To account for this, more sophisticated neutron experiments are needed \cite{brownJPCS}. We suggest that the non-collinear part of the magnetization density can probably be  extracted by combining flipping ratio and integrated intensity measurement in a magnetic field, but this needs to be verified  experimentally.

A similar maximum entropy reconstruction (using a trigonal symmetry subgroup of {\it Fd$\bar{3}$m}) was performed on flipping ratios measured with the field parallel to [111]. The resulting  magnetization density component $m_Z$ obtained at 10K in 1 Tesla  is shown in Fig.\ref{fig2}. It gives an  evidence that  Tb ions having  their local axis parallel to the field have a very high magnetic moment, while the  rest of the Tb ions  have very small moments, in agreement with the biased "one in-three out" structure obtained by the refinement (bottom  Fig.\ref{fig1} a).
\begin{figure}[t]
\begin{center}
\includegraphics[width=16pc]{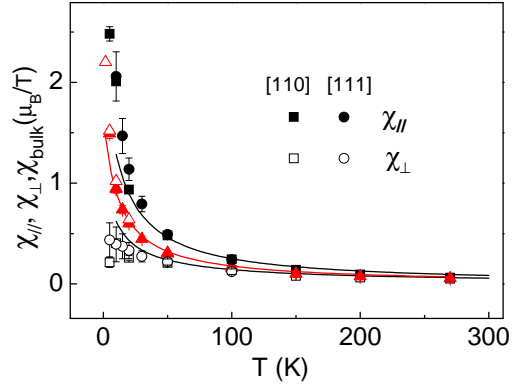}
\end{center}
\caption{\label{fig3}Longitudinal and transverse local susceptibilities
 $\chi_\parallel$ and  $\chi_\perp$  versus T. Square and circle symbols correspond to a field along [110] and [111] respectively. Filled  triangles show the bulk susceptibility derived from the neutron data, $\chi^{bulk}_n$=1/3 $\chi_\parallel$ +2/3 $\chi_\perp$, open triangles
the bulk susceptibility from Ref. \cite{Mirebeau07}. Lines are calculations using the
CF parameters of \tbti\ (see text).}
\end{figure}

\section{Longitudinal and transverse local susceptibilities}
 Figure \ref{fig3} shows
that the ratio of the main axes of the ellipsoids $\chi_\parallel$/$\chi_\perp$ increases with
decreasing temperature. This reflects the increasing CF anisotropy as temperature decreases. Indeed, at high temperature, many CF states are populated and the anisotropy is close to that of the free ion (weak or Heisenberg-like), whereas at low temperature, the lowest doublets alone are populated, which have a
strong Ising behaviour.
The local susceptibilities obtained with the field along [110] and [111] are in a very good agreement,
which confirms that this approach is valid regardless the field orientation.

For comparison with the experimental data, we used the
CF parameters of \tbti\ derived from the inelastic neutron spectra \cite{Mirebeau07}. The parameters account properly for the  trigonal symmetry of the Tb environment and allow
to compute the  magnetization induced by a magnetic field applied parallel
or perpendicular to the local $<111>$  axis.
Interatomic exchange and dipolar interactions were taken into account in
the molecular field approximation via a microscopic constant $\lambda$.
The self-consistent calculation, valid
down to 10 K, yields a constant $\lambda$ = -0.35 T/$\mu_B$, close to the
value derived in Ref. \cite{Mirebeau07}. The result of the calculations agrees well with the experiment, see Fig.\ref{fig3}.

In the local susceptibility approach, the atomic site susceptibility
tensor accounts only for the linear paramagnetic response
of the moments to an applied field. The range of fields and temperatures satisfying the  condition of  linear response can be estimated from the bulk magnetization data \cite{Mirebeau07}: the $\chi_{ij}$ parameters obtained in our refinement allow the general behavior of \tbti\ to be described above 10 K in fields up to 7 T applied in an arbitrary  direction.

\section{Conclusion}
 We determined the field induced magnetic structures in \tbti\ by polarized
neutron diffraction, in a wide range of temperatures and fields, applied in the [110] or [111] directions.
The [110] low temperature ground state shows a non-collinear
magnetic structure that resembles the local spin ice ground configuration, but
with Tb moments of very different magnitudes. The [111] low temperature ground state is close to the structure "one in-three out" found in spin ices, with the difference that all Tb  moments have an additional  component along the [111] direction due to the magnetic field. This is attributed to the finite CF anisotropy of \tbti\ which is much weaker than in model spin ices.
Polarized neutron diffraction
and the local susceptibility approach provide an easy way for
studying field induced structures in pyrochlore compounds. They provide  a universal description of the magnetic structures in \tbti\ in a large range of fields and temperatures, with only two local susceptibility parameters regardless of the field direction.

We thank P. J. Brown for help in using CCSL and interesting
discussions and G. Dhalenne for the crystal synthesis. H. Cao acknowledges support from the Triangle de la
Physique.

\end{document}